# Strategies for optimizing plasmonic grating couplers with topology-based inverse design

Michael Efseaff[1] and Mark C. Harrison[1,*]

[1]*Chapman University, Fowler School of Engineering, 1 University Dr., Orange, CA 92866, USA*
*\*mharrison@chapman.edu*

**Abstract:** Numerical simulations have become a cornerstone technology in the development of nanophotonic devices. Specifically, 3D finite-difference time domain (FDTD) simulations are a widely used due to their flexibility and powerful design capabilities. More recently, FDTD simulations in conjunction with a design methodology called inverse design has become a popular way to optimize device topology, reducing a device's footprint and increasing performance. We implement a commercial inverse design tool to generate complex grating couplers and explore a variety of grating coupler design methodologies. We compare the conventionally designed grating couplers to those generated by the inverse design tool. Finally, we discuss the limitations of the inverse design tool and how different design strategies for grating couplers affect inverse design performance, both in terms of computational cost and performance of the resulting device.

## 1. Introduction

No matter how efficient individual devices in a photonic circuit are, the circuit can only perform properly if a sufficient amount of power is coupled into the devices that make it up. Coupling has become increasingly important as photonic technology moves towards commercialization and reliably packaging photonic integrated circuits is a key challenge to overcome [1–4]. Getting the light signal to the circuit is typically done with either grating couplers or end fire (edge) couplers [5–8]. Both have their advantages and disadvantages and are often used for specific applications [9]. However, as simulation and optimization tools continue to advance, grating couplers seem to benefit more than end fire couplers from advanced optimization and design methods [5,10–12].

Grating couplers are a cornerstone device and are one of the main ways signals can enter and exit photonic integrated circuits. As such, making a highly efficient and compact grating coupler is extremely important to the performance of an entire circuit [13]. One of the most significant advantages of grating couplers is that they can be placed anywhere on a chip to allow a signal to enter due to their vertical coupling. End fire couplers, while typically having higher coupling efficiency, only allow for the signals to be coupled in horizontally. In other words, end fire couplers are limited to being placed on the edges of the chip. This makes grating couplers more flexible in terms of circuit layout.

The flexibility grating couplers provide is only as impactful as the device's overall efficiency. Powerful simulation tools, including finite-difference time domain (FDTD) simulations, have allowed researchers and engineers to quickly iterate on efficient grating coupler designs. As researchers explore using techniques such as inverse design and machine learning to optimize coupling efficiency, the potential approaches for optimization have ballooned [14–17]. Inverse design, in particular, has become a popular approach, but there are still many different ways to implement inverse design with regards to grating couplers. Grating couplers are, by nature, very sensitive to parameter shifts and the slightest changes in the grating period or duty cycle can change the expected performance of the device. Inverse design can help this issue because the device can be optimized over a range of wavelengths and the structures generated consider over and under etching during fabrication. Therefore, inverse design allows us to make more efficient devices that are also more tolerant to manufacturing errors.

Plasmonic grating couplers are a less common approach compared to the simpler design of fully dielectric photonic grating couplers [18,19]. However, the use of plasmonics for integrated photonic devices allows for some powerful advantages. The main draw to using plasmonics over silicon photonics is the smaller footprints required for plasmonic devices [20–22]. Plasmonics allows light to be confined to a much smaller area than that of silicon photonics, as such we can design plasmonic grating couplers that are a fraction of the size of their purely dielectric silicon photonic counterparts [23]. However, even in these cases, grating coupler sizes are directly tied to the overall spot size of the source, so with a large spot size the grating coupler must be large as well to capture the light. Plasmonic designs can be paired with other techniques to limit the input spot size for efficient and compact grating couplers. However, the major downside to using plasmonics is the high loss compared to purely dielectric photonic devices. The loss can be reduced by using long-range or dielectric-loaded surface-plasmon polariton (DLSPP) waveguides, which have inherently lower loss [24,25]. Combining inverse design with plasmonics offers the promise of small grating couplers that have lower loss than a conventionally designed plasmonic grating.

We explore different approaches to the design of compact, efficient, plasmonic grating couplers using inverse design to optimize transmission of light signals into photonic circuits. To ensure our designs are as efficient as possible we use an inverse design tool packaged with Ansys Lumerical. The inverse design tool we implement is an optimization method that uses gradient descent and the adjoint method to design an optimized device shape to maximize the figure of merit, which in our case is the device output [26]. Specifically, we implement a topological optimization which designs an overall device topology. This approach can make the optimizations take longer, but typically yields better results than other inverse design approaches. To keep our plasmonic grating couplers compact, we use a lensed fiber which focuses the light into a much smaller spot size compared to a typical fiber. Additionally, by using a plasmonic grating coupler we can avoid mode conversion that would be required if we were to use a purely dielectric grating coupler with other plasmonic devices, such as our own previously designed devices [27]. Designing efficient grating couplers that match the fabrication methods of other devices on-chip (such as using a DLSPP architecture for both) can be beneficial for the overall circuit design, even if the grating couplers are not the most efficient design possible when taking into account other fabrication methods. We find that the initial conditions of the inverse design are important to ensure it completes successfully, but do not necessarily alter the efficiency of resulting inverse-designed devices. Initial conditions do have an effect on how long inverse design optimizations take, but good conventional design can rival inverse design in efficiency for our plasmonic grating couplers. The initial conditions can also have an impact on other performance metrics, such as the shape of the transmission spectrum and bandwidth of the device. We explore various pros and cons of using inverse design based on comparing important metrics between our conventional and inverse designs.

## 2. Methods

We used Ansys Lumerical's FDTD simulation software along with its built-in inverse design tool, LumOpt, to complete this work [28]. Because we did not design this inverse design tool, we do not go into detail about its working principles here. The topology was optimized on a grid with 20 nm by 20 nm cells, and we used a radius filter set to 500 nm to enforce a minimum feature size on the resulting designs. Compared to other devices we have optimized with Lumerical's inverse design tool, the grating couplers have a vastly more costly computation time. The grating couplers are much larger and more complex than the other devices we have optimized, mainly due to the overall footprint of the devices and the inclusion of the vertical fiber. Because of the vertical fiber, the simulation region is about twice the volume of many other device simulations, which can make use of compact mode sources. To reduce the simulation size, a typical approach is to use a 2D side view (vertical cross-section) optimization for grating couplers [11,29]. This approach only allows one to optimize the spacing and size or shape of straight teeth, not the topology of the device. Therefore, it is most useful for gratings that have large widths in the dimension that is not simulated. In contrast, we performed full 3D simulations. In order to arbitrarily optimize the topology of a grating coupler and take into account the interactions of plasmonics at the metal interface, we need to do a full 3D simulation. These simulations are much more computationally intensive than 2D simulations and lead to much longer times for the simulations (and therefore the optimization process) to complete. Unfortunately, this 3D approach initially led to simulations were so complex that the inverse design was causing our server to crash, resulting in the inverse design taking potentially months to finish. Therefore, we took a modular approach to running the inverse design with the grating coupler when necessary. We decided to split the simulation into two parts: first we optimized the grating coupler teeth region and then optimized the taper region. While this methodology potentially results in slightly worse overall results compared to optimizing the entire grating coupler at once, the time to completion for the optimization drops to just weeks instead of months, which is a worthwhile tradeoff for us. For the design with a better initial condition, we found that splitting the optimization region into two parts was unnecessary for the simulations to complete in a reasonable time.

Another improvement to the overall run time of the simulations came from the amount of sampling done when recording the transmission at the output waveguide. When we increased the amount of wavelength data points recorded by the figure of merit monitor, the simulation time for a standard conventional simulation was reduced from hours to minutes. While it may sound unintuitive that allowing Lumerical to record more data allows the simulations to go faster, the larger amount of sampling points helps Lumerical's sampling algorithm interpret the simulation results quicker. Lumerical uses a Fourier transform algorithm and normalization to convert its time-domain simulations to frequency data, which evidently benefits from higher sampling rates. We found that a satisfactory number of data points to records at the output waveguide was 50, which was also recommended by Lumerical. Therefore, 50 data points was used for all the simulations.

For many different inverse design tools, the resulting designs and performance depend on the initial conditions used for the optimization [15,16,30]. The specific designs produced by the inverse design tool we use are also dependent on the initial conditions when starting the optimization. This is because there are a huge number of degrees of freedom for the device being produced by the inverse design algorithm for these large topology optimizations. For one of our inverse designs with 20 nm x 20 nm cells and an overall size of 15.4 μm by 10 μm for the optimization region, there are 385,000 total cells, each of which is a degree of freedom in our design. Due to the long duration for the inverse design to complete, we must ensure that running the inverse design is worth the time that it takes. Different initial conditions typically result in different design topologies, but they may result in similar device performance, as defined by the figure of merit. The adjoint method inverse design algorithm we use calculates the next step for the device topology using gradient descent, meaning the inverse design can get stuck at a local minimum because of the initial condition. A better initial condition might allow the inverse design to reach the global minimum, which is our goal in exploring different initial conditions. Because of the large number of degrees of freedom, it is likely that it will end up in a different local minimum. Two designs resulting from different initial conditions may have similar figures of merit, but different performance for other metrics that are not explicitly optimized. Therefore, we explored two major initial designs for the grating coupler. The first is a basic design with straight rectangles of dielectric material as the teeth and a small triangular taper region. The potential issue with this design is that the small taper region could lead to large loss since the light is being funneled to the output so quickly, which doesn't allow for adiabatic mode conversion. The other design we explored was similar to the previous design, however the teeth are curved. This allowed the teeth to act as a more gradual taper to allow the light to funnel into the output waveguides more adiabatically, at the expense of being larger. Both of these designs follow a DLSPP waveguide set up as shown in Fig. 1.

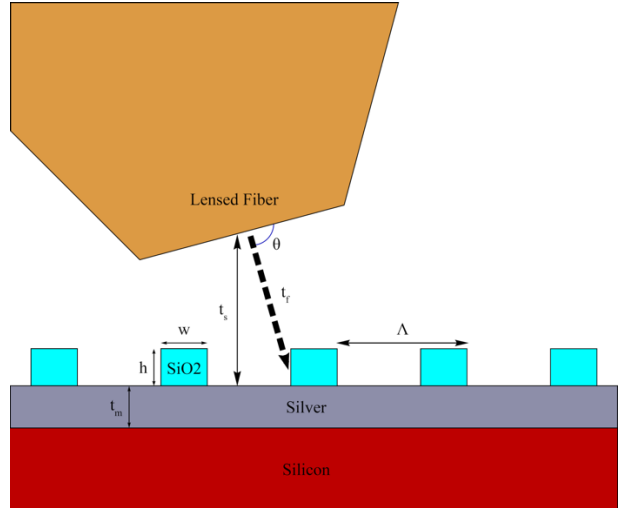

Fig. 1. General set up for all our grating coupler simulation where Λ = 1.54 μm, θ = 15°, t_f (focal length) = 1.44 μm, t_s (source height) = 1.66 μm, w = 520 nm, h = 440 nm, t_m (metal thickness) = 500 nm.

There are two main methods we used for generating the conventional grating coupler designs. The first method was using the grating period (Λ) defined in Eq. (1) to generate the conventional design, with the rest of the parameters (duty cycle, waveguide dimensions, etc.) being adjusted by hand to improve performance. These parameters were adjusted using simulations and simple parameter sweeps.

$$\Lambda = \frac{q\lambda}{n_{eff} - n_c \sin(\theta)} \quad (1)$$

In this equation, q is a diffraction order (which we set to 1), and $n_c$ is the refractive index of the cladding material (in our case the refractive index of air, 1), λ is the free-space wavelength (1310 nm), and θ is the angle of incidence of light on the grating (15°). We calculated $n_{eff}$ using Lumerical's mode source object to calculate the mode at the output waveguide and give us an effective index ($n_{eff}$) of 1.18. The grating period was calculated to be 1.54 μm. We used this calculation to generate a straight tooth grating coupler as well as a curved tooth grating coupler with ellipsoidal teeth. However, for the curved grating coupler design specifically, there are some secondary parameters present that can greatly affect the performance of the device. Because of this, we adapted an equation originally used to generate the structure for a purely dielectric grating coupler given in Eq. (2) [31].

$$\frac{\left(x + \frac{q\lambda n_1 \sin(\theta)}{n_{eff}^2 - n_c^2 \sin(\theta)^2}\right)^2}{\left(\frac{q\lambda n_{eff}}{n_{eff}^2 - n_c^2 \sin(\theta)^2}\right)^2} + \frac{y^2}{\left[\frac{q\lambda}{(n_{eff}^2 - n_c^2 \sin(\theta)^2)^{1/2}}\right]^2} = 1 \quad (2)$$

This equation gives the lengths for the major and minor axes for ellipses which are the teeth of the curved grating coupler, all sharing a focal point at the origin. The coordinates are given by x and y. In this equation, q represents which tooth is being generated, $n_c$ is the refractive index of the cladding material (air), λ is the free-space wavelength (1310 nm), and θ is the angle of incidence of light on the grating (15°). By changing the coefficient q, we can incrementally generate the axis lengths for subsequent teeth. This equation provided more details for creating the conventional design, but we still needed to manually adjust the secondary parameters such as duty cycle and tooth size. Once again, we used simulations and simple parameter sweeps to tweak these secondary parameters for better performance. Using these methods, we tested a straight tooth grating coupler using Eq. (1), and an elliptical curved tooth grating coupler using Eq. (2).

When running the inverse design, typically the optimization region matches the overall effective device area, but can produce designs smaller than the initial condition. Because of the behavior of grating couplers, we found that for the fully 3D optimizations we implemented with Lumerical's inverse design tool, we must use the conventional topology as the initial condition for the inverse design. In 2D side view cases, this is often not the case. It is also not the case for a recent work optimizing multilayer silicon photonic grating couplers in 3D [30]. Our need to use conventional topologies for initial conditions may have to do with the inverse design tool we are using or because we are optimizing plasmonic structures. Due to this constraint, the optimization region will at least be the same size or even bigger than the conventional design. By allowing for the optimization area to be bigger than the device region, we can allow for the inverse design to generate its own teeth structures that expand beyond the design and size of the original conventional grating coupler. We explore both options for the optimization region.

We tested variable optimization region sizes by taking our conventional straight tooth grating coupler and running the inverse design twice. The first inverse design had an optimization region that was the same as the conventional device size at 73.67 $\mu m^2$, and the second had an optimization region that was significantly larger at 129 $\mu m^2$. After running the simulations, we found that the inverse design that matched the conventional device size performed better than the larger area. This was surprising to us as we expected having a larger space would mean more design flexibility for the algorithm and better coupling. However, we believe having the parameter space larger than the conventional device provided the inverse design with more ways to get stuck at local minima, resulting in a worse overall design. So, for the rest of our inverse designs, we kept the optimization region sizes roughly the same as the conventional device area.

The simulation light source is extremely important because the gratings size shape and general functionality is dependent on the various parameters of the light source, such as angle of incidence, waist radius, wavelength, and distance from the device. For our purposes, we imitated our experimental set up, which in turn emulates the conditions of a device when it is fabricated and packaged for use. This setup consists of a lensed fiber directly above the grating 1.44 µm above the midpoint of the dielectric material, at an angle of incidence of 15°. The lensed fiber has a spot size of 2 µm at its focal point, which we imitate in the simulation. Using a lensed fiber, we can focus down the light before it reaches the grating, allowing us to reduce the size of the device, as opposed to a standard fiber where the spot size on the grating is much larger. However, even with reduced grating sizes, it can be beneficial to allow enough horizontal size to adiabatically convert the mode from the fiber to the waveguide mode. In the case of the curved-tooth grating coupler, the larger size also arose from the equations used to generate the shape of the teeth. Additionally, using the lensed fiber with a slightly over-sized grating coupler gives us tolerance to slight misalignment in an experimental setting.

## 3. Results

We compare our designs by measuring insertion loss, coupling efficiency, area, time to completion, and bandwidth. The insertion loss is defined in Eq. 3.

$$loss[dB] = -10\log\left(\frac{P_{out}}{P_{in}}\right) \quad (3)$$

In this equation, $P_{out}$ is the power coupled into the output waveguide and $P_{in}$ is the input power to the device. Coupling efficiency is a percentage representation of how much of the light coupled into the grating coupler. The area was calculated as the effective device region of the taper and teeth regions. The time to completion was calculated by taking the total time of the entire inverse design to show how simulation size and strategy affect the computational cost of the simulation. It is important to note that simulation time can vary greatly based on the computing resources used to complete the optimizations. However, by providing times for both conventional and inverse designs on the same equipment, we can give a sense of how much longer the inverse design takes than conventional design methods. The bandwidth of our devices is calculated using the full-width half-maximum approach where we say the bandwidth is equivalent to the range of wavelengths that allow for half or more of the transmission of the peak wavelength.

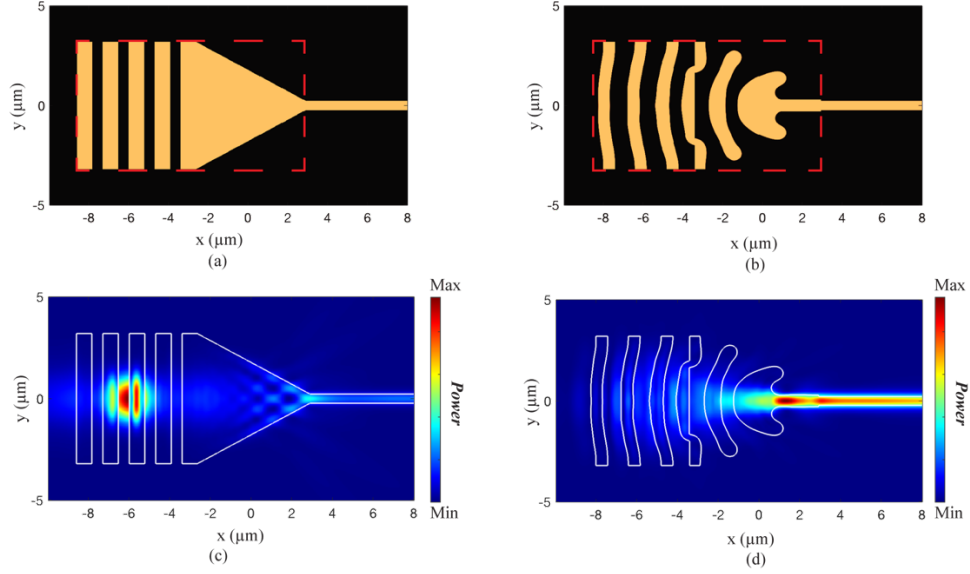

Fig. 2. Simulation results for the straight tooth grating couplers. (a) Conventional designed grating coupler topology. (b) Inverse designed grating coupler topology. The red dashed line in (a) and (b) indicates the area of the grating coupler used for calculation. (c) Conventional designed grating coupler power. (d) Inverse designed grating coupler power.

The topologies for the conventional and generated inverse design straight tooth grating couplers are shown in Fig. 2(a) and Fig. 2(b) respectively. The power inside the devices in a horizontal cross-section is depicted in Fig. 2(c) and Fig. 2(d). For this design, the optimization region was split into two parts with no overlap after the last gap in the conventional device, at a position of approximately x = -3.5. In the inverse design shown in Fig. 2(b) and Fig. 2(d), perfectly vertical features and a slight discontinuity is visible where the optimization region was split. Overlapping the split optimization regions could alleviate this vertical discontinuity, as could placing it in an area where a natural gap exists in the conventional grating coupler design. We can see that the topology of the inverse designed grating coupler deviates heavily from the conventional, with several unintuitive features. For example, the output waveguide in the optimization region is larger than the waveguide outside of the region, which we have observed before with this inverse design tool. The collective result of the new inverse design topology is that it allows for a greater amount of power to be transferred to the output waveguide. It is important to note that Fig. 2(c) and Fig. 2(d) do not have the same scale for power. If they had the same scale, then the power in Fig. 2(c) would be almost indistinguishable, because the grating coupler is much less efficient. This difference in scale and efficiency is the reason Fig. 2(c) shows a clear input spot from the light source. The other grating couplers redirect the light more efficiently, even though all simulations were excited by a source with the same parameters. We calculate that the insertion loss, coupling efficiency, and bandwidth are all significantly better in the inverse design than the conventional as seen in Table 1. Note that the time to completion only accounts for the actual duration of the simulations recorded by Lumerical, not including any manual adjustments, which partially contributes to the much shorter time for conventional designs.

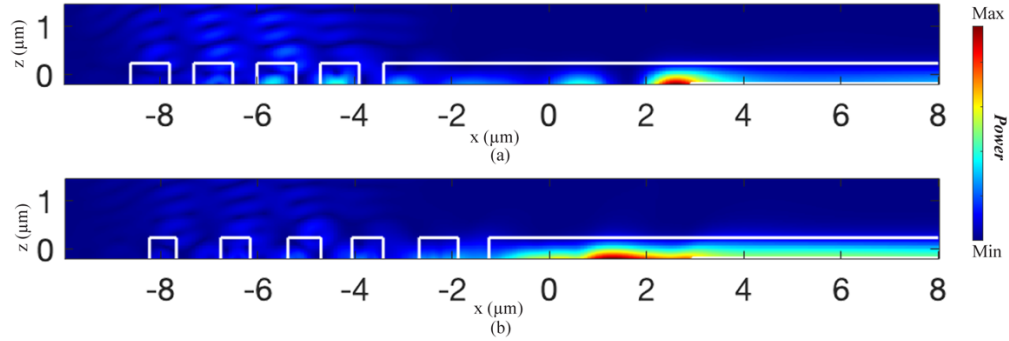

Fig. 3. Simulation results showing power for the straight tooth grating coupler at a vertical cross-section. (a) Conventional design and (b) inverse design

**Table 1. Straight Tooth Grating Coupler Characteristics**

| Device Design | Insertion Loss | Coupling Efficiency | Bandwidth | Area | Time to Completion (DD:HH:MM) |
|---|---|---|---|---|---|
| Conventional | 13.37 dB | 4.6% | 1175 nm - 1340 nm | 73.67 μm² | 00:00:28 |
| Inverse | 9.25 dB | 11.9% | 1215 nm - 1370 nm | 73.67 μm² | 39:19:59 |

Fig. 3(a) and Fig. 3(b) depict a centered vertical slice of the grating coupler to visualize loss from vertical reflections. From the figure we can see that the conventional design has more vertical reflections at the teeth, relative to the power within the taper and waveguide, compared to the inverse design.

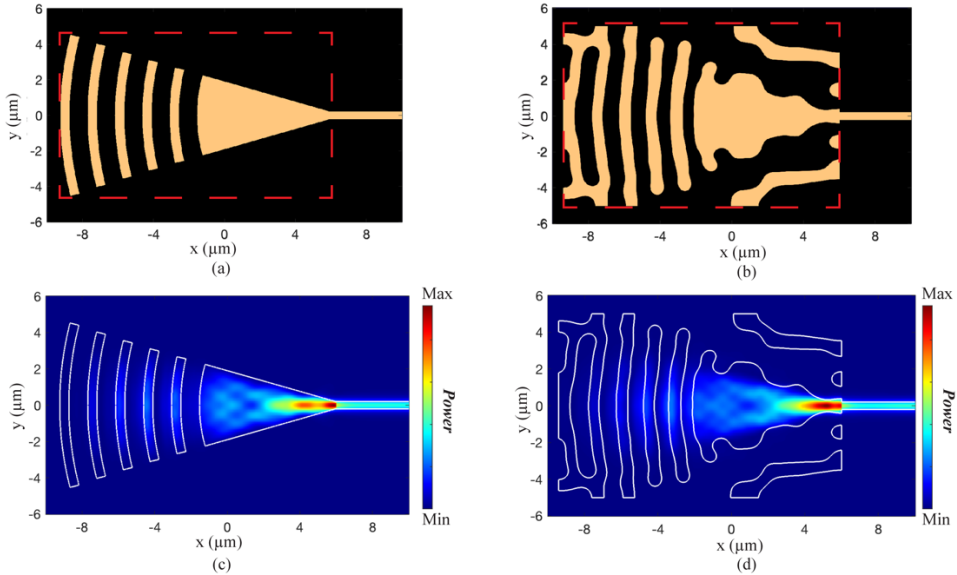

Fig. 4. Simulation results for the curved tooth grating couplers. (a) Conventional designed grating coupler topology. (b) Inverse designed grating coupler topology. The red dashed line in

(a) and (b) indicates the area of the grating coupler used for calculation. (c) Conventional designed grating coupler power. (d) Inverse designed grating coupler power.

The topologies for the conventional and inverse design for the curved tooth grating couplers are shown in Fig. 4(a) and Fig. 4(b) respectively. The power inside the devices in a horizontal cross-section is depicted in Fig. 4(c) and Fig. 4(d). For this design, we did not need to split the optimization region into two parts. We see that the topology of the inverse designed grating coupler deviates heavily from the conventional in Fig. 4, but still shares the curved design. In this design we also see some unintuitive features typical of inverse design such as a mismatch in output waveguide size and additional structures outside of the taper region. Once again, Fig. 4(c) and Fig. 4(d) do not have the same scale for power. However, due to similarity in efficiency, their scales are very close. This is why a clear input spot from the light source is not visible like it is in Fig. 2(c). We calculate that the insertion loss, coupling efficiency, and bandwidth are quite similar between the inverse design and the conventional as seen in Table 2.

Fig. 5(a) and Fig. 5(b) depict a centered vertical slice of the curved tooth grating coupler to visualize loss from vertical reflections. From the figure we can see that both the conventional and inverse designed structures have a very similar amount of vertically reflected power relative to the power within the taper and waveguide.

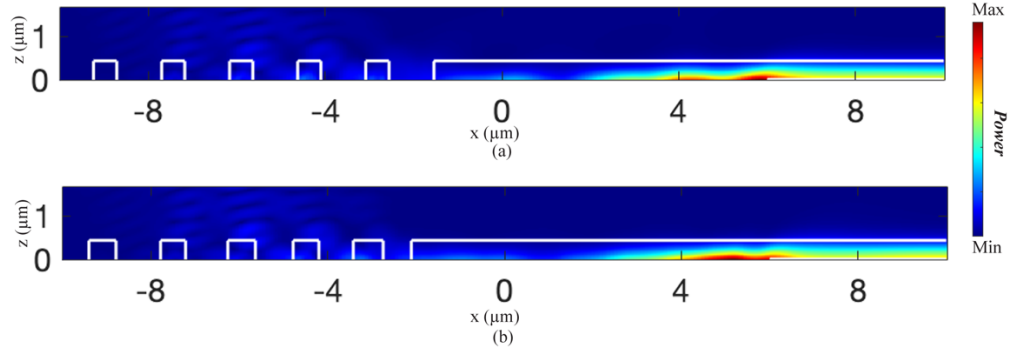

Fig. 5. Simulation results showing power for the curved tooth grating coupler at a vertical cross-section. (a) Conventional design and (b) inverse design

Table 2. Curved Tooth Grating Coupler Characteristics

| Device Design | Insertion Loss | Coupling Efficiency | Bandwidth | Area | Time to Completion (DD:HH:MM) |
| --- | --- | --- | --- | --- | --- |
| Conventional | 9.38 dB | 11.5% | 1210 nm - 1395 nm | 144.6 µm² | 00:00:24 |
| Inverse | 9.32 dB | 11.7% | 1233 nm - 1393 nm | 154 µm² | 20:02:29 |

Fig. 6 depicts the transmission spectra for all four of the designed devices. We can see that all devices except for the conventional straight tooth waveguide coupler have transmission spectra roughly centered at our target wavelength of 1310 nm. These simulations also have a similar transmission percentage at the peak. In contrast, the conventional straight tooth waveguide coupler is substantially worse than the other designs, with a spectrum that is off-center from the target wavelength and a substantially lower transmission peak percentage.

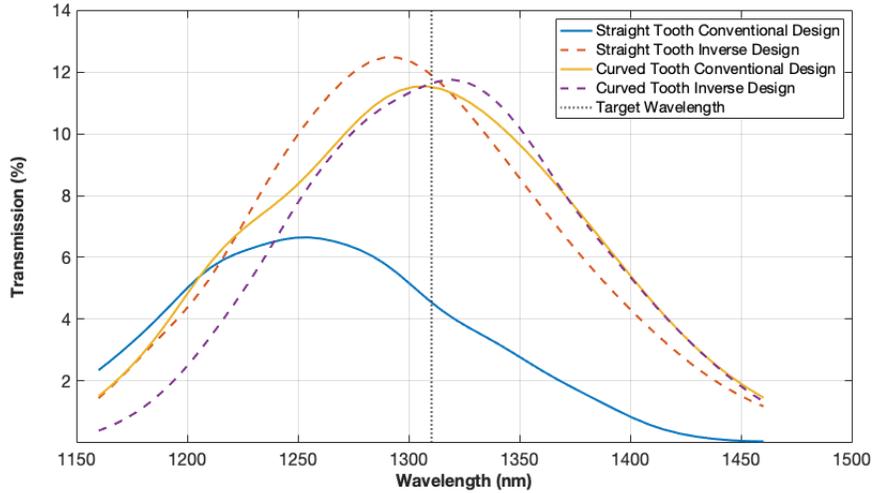

Fig. 6. Transmission spectra for all grating coupler designs, measured by how much power coupled into the output waveguide. The target wavelength of 1310 nm is indicated on the graph by a vertical dashed line.

## 4. Discussion

For the straight tooth grating coupler design the inverse design was able to improve the coupling efficiency by more than double. The bandwidths for the conventional and inverse design are relatively similar with the conventional design being slightly larger. However, the conventional straight tooth grating coupler has a transmission spectrum greatly off center from our target wavelength and the wide bandwidth is potentially just a result of the overall lack of transmission. Additionally, the spectrum could probably be made more centered by using some more advanced design techniques (such as apodization), which we ignored as the conventional design was primarily a starting point for the inverse design. In contrast, the inverse designed grating coupler's transmission spectrum has a defined peak close to our target wavelength of 1310 nm and has a more symmetrical gaussian shape. The spectrum is still skewed off-center slightly, likely due to the initial condition of the inverse design. The figure of merit we used for optimization was the overall transmission to the waveguide, and we did not optimize for a center wavelength. In addition, the inverse designed grating coupler has a narrower bandwidth since it was optimized for a narrow range of wavelengths between 1300 nm and 1320 nm. Furthermore, the inverse design added some curvature and slight apodization to the straight teeth which seems to help taper down the light signal more gradually and efficiently. We therefore predicted that if we started from a curved tooth or apodized design that the inverse design would perform better since the initial condition would be closer to an optimal design. We chose to investigate a curved tooth initial condition.

From a relative comparison standpoint, the inverse design for the straight tooth design performed much better than the conventional design, however the overall loss and coupling efficiency is still rather low compared to other forms of grating couplers. The grating coupler as seen in *Koev, et. al.* [18], while much larger than ours (40 μm x 30 μm), reached peak efficiencies of 45% coupling efficiency which is considerably higher than our devices. This discrepancy is likely due to a design that utilized grooved metal for grating teeth. One big limitation of our design is having a flat silver layer where the source is shining down directly on top of it. The metal teeth will interact with light more strongly than our dielectric teeth and will be better at redirecting the energy into the waveguide plane. This is also a limitation of the Lumerical inverse design tool itself. Lumerical can only optimize one dielectric material during the optimization, so there was no way for the tool to reshape the silver in some way to reduce the vertical reflections. We could have added in etches in the silver, similar to the previously discussed grating couplers, but this also results in a more complicated fabrication process.

The changes in the inverse design for the curved tooth grating coupler design were very different compared to the straight tooth grating. Both the inverse design and the conventional design for the curved tooth grating coupler had essentially the same coupling efficiency with the conventional design having a marginally larger bandwidth. In this case, the inverse design spectrum is slightly off-center compared to the conventional design. Once again, this is likely due to the fact that we did not optimize for a center wavelength in the inverse design, but just for overall transmission.

The similarity in efficiency between conventional and inverse designs for the curved tooth grating coupler contradicted our prediction that starting from a more efficient conventional design would produce a more efficient inverse design. The lack of improvement in the inverse design is most likely due to factors outside of the control of the inverse design tool. As previously mentioned, LumOpt, our inverse design tool, can only optimize one non-metallic material. This means that if there are issues with a device design that do not pertain to the shape of the optimized material, the inverse design will be greatly limited in what it can achieve. This is true for our grating coupler designs where the best designs had about 73% of the light from the lensed fiber reflected back upward, and the worst grating coupler design had about 81% of the light was reflected. A plasmonic grating coupler with only dielectric teeth will be inherently limited in how much light it can redirect, since the metal layer will interact with the light more strongly. Thus, no matter how efficient the inverse design is, it cannot completely overcome the physical limitations of the device architecture we choose. While it is possible for the inverse design to allow for less light to be reflected, as we see in the inverse designed straight toothed grating coupler, inverse design is not able to fix any inherent limitations of a design outside of its control. If we had considered etching the metal, we would most likely find much more efficient results like we see in *Ayata, et al* [23]. From this we can conclude that inverse design, like anything else, is not a catch-all solution for creating devices.

To achieve better coupling efficiencies, the variables outside of the inverse design tool's control need to be addressed. In the case of the grating coupler, it would be better to stick with a conventional design approach where modifications such as a change in fiber tilt angle can be accounted for easily to update the design. This is reinforced by how long the inverse design took. For essentially the same efficiency we had to wait weeks for the inverse design to finish, whereas in that time we could have been fine tuning a conventional design to potentially surpass what we have now. In contrast, other types of devices do benefit greatly from inverse design because we are primarily optimizing for size and loss.

One of the more interesting developments with the inverse designs for both the straight and curved teeth grating couplers were the similarities between them. Both optimizations favored creating curved teeth that had a slight parabolic shape to them. This is interesting considering the vast difference in initial condition between the two designs. It was surprising to us that for the curved initial condition the teeth straightened out to have a slight bend more similar to the inverse design of the straight teeth grating coupler. This is possibly due to the fact that since the design is trying to maximize the power throughput for a specific range of wavelengths, a gentler curve might be better, leading to similar results. The similarity between inverse designed device efficiencies indicates that when using inverse design tools there is limited utility to fully optimizing the design used for the initial condition. For example, the conventional design could be improved using an apodization of the grating period. Since both inverse designs produce a slight apodization for the grating teeth, it is likely not necessary to do so manually if one is using an inverse design tool. However, this observation supports our conclusion that if coupling efficiency is the most important metric, then for certain designs a conventional design approach produces comparable results but has more flexibility, and is a better design method. This is true for the DLSPP grating coupler we considered. These conventional design approaches could also include more advanced design techniques than we considered here, such as the aforementioned apodization. In other instances where factors such as size or bandwidth are more important, then inverse design can still be a useful tool. This is evident in the size of the straight tooth inverse designed device, which is small but still has high efficiency.

Compared to other state-of-the-art, inverse designed grating couplers, the size of the designs we produce compare favorably. For example, *Sapra, et al* design a silicon photonic grating coupler that is 12 µm x 12 µm excluding the taper [11], *Hammond, et al* design a multilayer silicon photonic grating coupler that is 10 µm x 10 µm for the entire structure [30], and *Andkjær, et al* design a 2.5 µm long plasmonic grating coupler, but do not report the width or comment on a taper [5]. The conventionally designed DLSPP grating coupler by *Ayata, et al* had dimensions of 12µm x 13.5 µm [23]. Our inverse designed device based on the straight tooth grating coupler is fairly small, measuring 6.4 µm x 11.52 µm. The curved-tooth grating coupler design was larger, at 10 µm x 15.4 µm, making it more similar in size to devices in other works.

Although we only performed simulations for this study, we expect that a similar performance can be achieved experimentally. As previously mentioned, we designed the devices using a light source that mimics our experimental setup. Furthermore, the inverse designed devices were optimized with a radius filter of 500 nm to ensure a minimum feature size that can be fabricated using common nanofabrication equipment, such as an e-beam writer. Unfortunately, the radius filter implemented in Lumerical's inverse design tool does not account for the edges of the optimization region. Because of this, both of our inverse designed devices have sharp edges with a radius less than 500 nm at the boundary, which would likely reduce experimental performance for these devices. DLSPP waveguides have been experimentally studied and others have made conventional designs similar to ours [23,32]. There might be some degradation in performance, as often happens experimentally, but most of the degradation is likely to be similar between the conventional and inverse designs. For example, in both cases, the designs have perfectly vertical sidewalls, which may be difficult to fabricate. Although it is possible to optimize for varying fabrication tolerances in an inverse design process, we did not do that in this study.

## 5. Conclusion

Using inverse design, we generated new designs for plasmonic grating couplers using different conventional designs for initial conditions. We compare the inverse designs to the conventional designs to explore the strengths and limitations of the inverse design tool. We reduced the size of the grating couplers, which is normally a disadvantage, by using a plasmonic structure and a lensed fiber to reduce the spot size of the input light beam. The next steps are to fabricate these devices and compare the simulation results against the experimental results with the physical devices. We could also work to improve performance by taking a hybrid approach to the grating coupler, using a highly efficient and standard purely dielectric grating for the coupling and then transitioning the signal to a plasmonic mode for coupling into other plasmonic devices. This would allow us to work around some of the limitations we have in our design structure and the Lumerical inverse design tool, while still preserving the advantages of plasmonics in other types of devices.

**Funding.** This material is based upon work supported by the Air Force Office of Scientific Research (Award No. FA9550-21-1-0188).

**Acknowledgments.** The authors would like to thank Scott Cummings for early preliminary work in creating conventional grating coupler simulations.

**Disclosures.** The authors declare no conflicts of interest.

**References**

1. S. Nambiar, P. Sethi, and S. Selvaraja, "Grating-Assisted Fiber to Chip Coupling for SOI Photonic Circuits," Applied Sciences **8**, 1142 (2018).
2. S. Khan, S. M. Buckley, J. Chiles, R. P. Mirin, S. W. Nam, and J. M. Shainline, "Low-loss, high-bandwidth fiber-to-chip coupling using capped adiabatic tapered fibers," APL Photonics **5**, 056101 (2020).
3. L. Brusberg, A. R. Zakharian, Ş. E. Kocabaş, L. W. Yeary, J. R. Grenier, C. C. Terwilliger, and R. A. Bellman, "Glass Substrate With Integrated Waveguides for Surface Mount Photonic Packaging," Journal of Lightwave Technology **39**, 912–919 (2021).
4. L. Carroll, J.-S. Lee, C. Scarcella, K. Gradkowski, M. Duperron, H. Lu, Y. Zhao, C. Eason, P. Morrissey, M. Rensing, S. Collins, H. Hwang, and P. O'Brien, "Photonic Packaging: Transforming Silicon Photonic Integrated Circuits into Photonic Devices," Applied Sciences **6**, 426 (2016).
5. J. Andkjær, S. Nishiwaki, T. Nomura, and O. Sigmund, "Topology optimization of grating couplers for the efficient excitation of surface plasmons," J. Opt. Soc. Am. B **27**, 1828 (2010).
6. L. Cheng, S. Mao, Z. Li, Y. Han, and H. Y. Fu, "Grating Couplers on Silicon Photonics: Design Principles, Emerging Trends and Practical Issues," 25 (2020).
7. C. Fisher, L. C. Botten, C. G. Poulton, R. C. McPhedran, and C. M. de Sterke, "End-fire coupling efficiencies of surface plasmons for silver, gold, and plasmonic nitride compounds," J. Opt. Soc. Am. B **33**, 1044 (2016).
8. Y. M. Morozov, A. S. Lapchuk, M.-L. Fu, A. A. Kryuchyn, H.-R. Huang, and Z.-C. Le, "Numerical analysis of end-fire coupling of surface plasmon polaritons in a metal-insulator-metal waveguide using a simple photoplastic connector," Photon. Res. **6**, 149 (2018).
9. R. Marchetti, C. Lacava, L. Carroll, K. Gradkowski, and P. Minzioni, "Coupling strategies for silicon photonics integrated chips [Invited]," Photon. Res. **7**, 201 (2019).
10. D. Melati, Y. Grinberg, M. Kamandar Dezfouli, S. Janz, P. Cheben, J. H. Schmid, A. Sánchez-Postigo, and D.-X. Xu, "Mapping the global design space of nanophotonic components using machine learning pattern recognition," Nat Commun **10**, 4775 (2019).
11. N. V. Sapra, D. Vercruysse, L. Su, K. Y. Yang, J. Skarda, A. Y. Piggott, and J. Vuckovic, "Inverse Design and Demonstration of Broadband Grating Couplers," IEEE J. Select. Topics Quantum Electron. **25**, 1–7 (2019).
12. C. Sideris, E. Garza, and O. P. Bruno, "Ultrafast Simulation and Optimization of Nanophotonic Devices with Integral Equation Methods," ACS Photonics **6**, 3233–3240 (2019).
13. S. Y. Siew, B. Li, F. Gao, H. Y. Zheng, W. Zhang, P. Guo, S. W. Xie, A. Song, B. Dong, L. W. Luo, C. Li, X. Luo, and G.-Q. Lo, "Review of Silicon Photonics Technology and Platform Development," J. Lightwave Technol. **39**, 4374–4389 (2021).
14. S. Hooten, R. G. Beausoleil, and T. Van Vaerenbergh, "Inverse design of grating couplers using the policy gradient method from reinforcement learning," Nanophotonics **10**, 3843–3856 (2021).
15. A. Michaels and E. Yablonovitch, "Inverse Design of Near Unity Efficiency Perfectly Vertical Grating Couplers," Opt. Express **26**, 4766 (2018).
16. L. Su, R. Trivedi, N. V. Sapra, A. Y. Piggott, D. Vercruysse, and J. Vučković, "Fully-automated optimization of grating couplers," Opt. Express, OE **26**, 4023–4034 (2018).
17. X. Tu, W. Xie, Z. Chen, M.-F. Ge, T. Huang, C. Song, and H. Y. Fu, "Analysis of Deep Neural Network Models for Inverse Design of Silicon Photonic Grating Coupler," Journal of Lightwave Technology **39**, 2790–2799 (2021).
18. S. T. Koev, A. Agrawal, H. J. Lezec, and V. A. Aksyuk, "An Efficient Large-Area Grating Coupler for Surface Plasmon Polaritons," Plasmonics **7**, 269–277 (2012).
19. Q. Gao, F. Ren, and A. X. Wang, "Direct and Efficient Optical Coupling Into Plasmonic Integrated Circuits From Optical Fibers," IEEE Photon. Technol. Lett. **28**, 1165–1168 (2016).
20. T. J. Davis, D. E. Gómez, and A. Roberts, "Plasmonic circuits for manipulating optical information," Nanophotonics **6**, (2017).
21. C. Hoessbacher, Y. Salamin, Y. Fedoryshyn, W. Heni, A. Josten, B. Baeuerle, C. Haffner, M. Zahner, H. Chen, D. L. Elder, S. Wehrli, D. Hillerkuss, D. Van Thourhout, J. Van Campenhout, L. R. Dalton, C. Hafner, and J. Leuthold, "Optical Interconnect with Densely Integrated Plasmonic Modulator and Germanium Photodetector Arrays," in *Optical Fiber Communication Conference* (OSA, 2016), p. Th1F.6.
22. M. R. Pav, N. Granpayeh, S. P. Hosseini, and A. Rahimzadegan, "Ultracompact double tunable two-channel plasmonic filter and 4-channel multi/demultiplexer design based on aperture-coupled plasmonic slot cavity," Optics Communications **437**, 285–289 (2019).
23. M. Ayata, Y. Fedoryshyn, U. Koch, and J. Leuthold, "Compact, ultra-broadband plasmonic grating couplers," Opt. Express **27**, 29719 (2019).
24. P. Berini, "Long-range surface plasmon polaritons," Adv. Opt. Photon. **1**, 484 (2009).
25. T. Holmgaard, J. Gosciniak, and S. I. Bozhevolnyi, "Long-range dielectric-loaded surface plasmon-polariton waveguides," Opt. Express **18**, 23009 (2010).

26. C. M. Lalau-Keraly, S. Bhargava, O. D. Miller, and E. Yablonovitch, "Adjoint shape optimization applied to electromagnetic design," Opt. Express **21**, 21693 (2013).
27. M. Efseaff, K. Wynne, K. Narayan, and M. C. Harrison, "Implementing commercial inverse design tools for compact, phase-encoded, plasmonic digital logic devices," Journal of Nanophotonics **17**, (2023).
28. "Photonic Inverse Design," https://www.lumerical.com/solutions/inverse-design/.
29. M. Khodami and P. Berini, "Grating couplers for (Bloch) long-range surface plasmons on metal stripe waveguides," J. Opt. Soc. Am. B **36**, 1921 (2019).
30. A. M. Hammond, J. B. Slaby, M. J. Probst, and S. E. Ralph, "Multi-layer inverse design of vertical grating couplers for high-density, commercial foundry interconnects," Opt. Express **30**, 31058 (2022).
31. R. Waldhäusl, B. Schnabel, P. Dannberg, E.-B. Kley, A. Bräuer, and W. Karthe, "Efficient Coupling into Polymer Waveguides by Gratings," Appl. Opt., AO **36**, 9383–9390 (1997).
32. M. G. Nielsen, J.-C. Weeber, K. Hassan, J. Fatome, C. Finot, S. Kaya, L. Markey, O. Albrektsen, S. I. Bozhevolnyi, G. Millot, and A. Dereux, "Grating Couplers for Fiber-to-Fiber Characterizations of Stand-Alone Dielectric Loaded Surface Plasmon Waveguide Components," J. Lightwave Technol. **30**, 3118–3125 (2012).